\documentclass[%
aip,
 reprint,
 amsmath,amssymb,
]{revtex4-1}

\usepackage{graphicx}
\usepackage{dcolumn}
\usepackage{bm}

\usepackage[utf8]{inputenc}
\usepackage[T1]{fontenc}
\usepackage{mathptmx}
\usepackage{etoolbox}

\usepackage{listings}

\usepackage{pgfplots}
\pgfplotsset{compat=1.15}

\usepackage{natbib}

\newcommand{\unit}[1]{\,\text{#1}}
\newcommand{\rnd}[1]{\left( #1 \right)}

\newcommand{\csat}{c_{\text{sat}}}
\newcommand{\rhoc}{\rho_{\text{c}}}

\usepackage{ulem}
\usepackage{xcolor, soul}
\sethlcolor{yellow}

\newcommand{\deleted}[1]{\sout{#1}}
\renewcommand{\deleted}[1]{}
\newcommand{\added}[1]{\emph{#1}}
\renewcommand{\added}[1]{#1}

\makeatletter
\def\@email#1#2{%
 \endgroup
 \patchcmd{\titleblock@produce}
  {\frontmatter@RRAPformat}
  {\frontmatter@RRAPformat{\produce@RRAP{*#1\href{mailto:#2}{#2}}}\frontmatter@RRAPformat}
  {}{}
}%
\makeatother

\begin{document}

\preprint{}

\title{\added{Uneven Extraction}\deleted{Dissolution Flow Instabilities} in Coffee Brewing}

\author{W. T. Lee}%
 \altaffiliation[Also at ]{MACSI, University of Limerick}

\author{A. Smith}%
\email{a.smith@hud.ac.uk}

\author{A. Arshad}%
\affiliation{%
  School of Computing and Engineering, University of Huddersfield, Queensgate, Huddersfield HD1 3DH, UK
}%

\date{\today}

\begin{abstract}
A recent experiment showed that, contrary to theoretical predictions, beyond a cutoff point grinding coffee more finely results in lower extraction. One potential explanation for this is that fine grinding promotes non-uniform extraction in the coffee bed. We investigate the possibility that this could occur due \deleted{to an instability involving} \added{the interaction between} dissolution and flow \added{promoting uneven extraction}. A low dimensional model in which there are two possible pathways for flow is derived and analysed. This model shows that, below a critical grind size, \deleted{the flow-dissolution instability contributes to} \added{there is }  a decreasing extraction with decreasing grind size \added{as is seen experimentally}.  \deleted{This strongly suggests that a flow instability can explain the observed decrease in extraction.}\added{In the model this is due to a complicated interplay between an initial imbalance in the porosities and permeabilities of the two pathways which is increased by flow and extraction, leading to the complete extraction of all soluble coffee from one pathway.}
\end{abstract}

\maketitle

\section{Introduction}

Espresso coffee is a beverage brewed from the roasted, ground cherries (beans) of the coffee robusta or arabica plant. In brewing an espresso hot (92-95$^\circ$C) water is forced at high pressure (9-10atmospheres) through a bed of 15-22g of finely ground coffee resulting in a beverage with a mass of 30-60g~\cite{cameron2020}. Although coffee is  a complex mix of nearly 2000 chemicals~\cite{Moroney2015} most mathematical models of coffee brewing treat coffee as a single substance using mass as a measure of the amount. The quality of coffee can be measured by two properties: strength and extraction yield. Strength is the mass concentration of dissolved coffee solids in the beverage. Extraction yield is the mass fraction of the coffee grains that have dissolved. Coffee grains are only partially soluble so there is a maximum value of the extraction yield which cannot be exceeded. A rough measure of coffee quality is given by the coffee quality control chart 
which plots strength against extraction yield. 

Fasano et al.~\cite{Fasano2000} developed a partial differential equation model of espresso brewing which treated coffee as a multicomponent substance  whilst Mo et al.\cite{mo2022modeling} considered the impact of particle swelling on extraction. However, the primary interest of the authors was in the mathematical properties of the equations e.g.\ investigating the existence and uniqueness of solutions rather than parameterising a model with experimental data. Moroney et al.~\cite{Moroney2015,Moroney2016,Moroney2019, boulais2020two} developed a multiscale model of cafetiere and filter coffee brewing based on models of groundwater flow and leaching. These models treated coffee as a dual porosity medium\cite{panfilov2019homogenized} with pores between grains of ground coffee and pores within ground coffee. The model also included a bimodal grain size distribution.  Cameron et al.~\cite{cameron2020} developed a model of espresso brewing to complement an experimental investigation of the effect of grain size on extraction \cite{antonopoulou2018numerical}. This model is based on mathematical models of lithium ion batteries. As with the models developed by Moroney et al. the model includes a bimodal grain size. A diffusion equation is used to model transport of coffee within grains to the intergranular pores. Smith and Lee~\cite{smith2021brewing} developed a simplified model of coffee brewing for teaching purposes. This was applied to cafetiere brewing and only considered a single grain size. 

Cameron et al.’s experiment looked at the trend of extraction yield with grind size parameterised by the setting of their grinder, denoted by $g$ here. Small values of $g$ indicate finely ground coffee, whilst larger values of $g$ result in more coarsely ground coffee.  Surprisingly a plot of extraction yield against grind size shows a peaked distribution with a maximum extraction at intermediate grind size values and less extraction seen for coarser and more finely ground coffee. This same trend of finer particles giving weaker coffee below a threshold value is also observed in other brewing methods~\cite{cordoba2019effect}. The expected trend was that extraction would increase as grind size decreased due to a combination of factors. These were: reduced permeability resulting in slower flow and thus liquid spending more time in contact with the coffee grains, a larger surface area over which transfer could occur and a smaller distance from the interior of the grains to the exterior. In order to explain the observed result, the authors hypothesised that at low grain sizes the entire coffee bed was not participating in extraction. In other words, there was an increasing trend of extraction with grind size, but this was counterbalanced by a reducing volume of the coffee bed available for extraction. The mathematical model developed in the paper initially predicted the expected trend of increasing extraction with decreasing grind size. In order to agree with the observed trend, a reduced area participating in extraction beyond the peak in extraction was imposed on the model. The authors speculated this might be due to fines clogging some pathways. 

Here we consider the possibility that an \deleted{instability linking} \added{interaction between} flow and extraction may explain the experimental data.  In  this hypothesis the fluid has access to every area of the coffee bed, but flow and thus extraction proceeds more slowly in some regions than others. This proposed \deleted{instability} \added{mechanism} is based on a positive feedback loop in which flow and extraction reinforce each other. Extraction causes an increase in porosity and thus permeability. This increase in permeability will in turn lead to more flow and so more extraction will occur. This suggests that small differences in porosity in the coffee bed will become amplified over time. Regions with higher porosity will see more flow and thus more extraction, whilst regions of lower porosity will see less flow and thus less extraction.  Usually up to around 30\% of the coffee is removed during the extraction \added{so there is scope for considerable variation in porosity and thus permeability}.

\section{Methods}

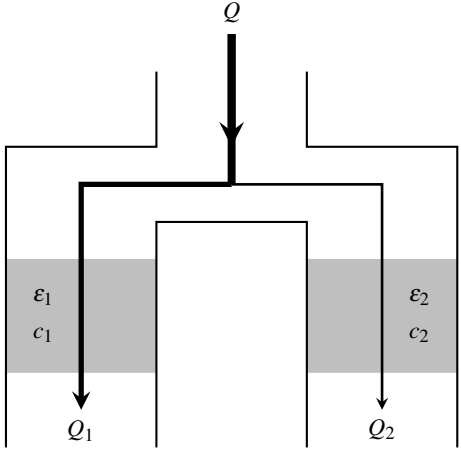
\begin{figure}
    \centering
\begin{tikzpicture}[]
  \draw[fill=lightgray,color=lightgray](0,2)--(0,3.5)--(2,3.5)--(2,2)-| cycle;
  \draw[fill=gray,color=lightgray](6,2)--(6,3.5)--(4,3.5)--(4,2)-| cycle;
  \draw[thick](0,1)--(0,5)--(2,5)--(2,6);
  \draw[thick](2,1)--(2,4)--(4,4)--(4,1);
  \draw[thick](6,1)--(6,5)--(4,5)--(4,6);
  \draw[line width=3pt,-stealth](3,6.5)node[above]{$Q$}--(3,5.0);
  \draw[line width=3pt](3,6.5)--(3,4.5);
  \draw[line width=2pt,-stealth](3,4.5)--(1,4.5)--(1,1.5)node[below]{$Q_1$};
  \draw[line width=1pt,-stealth](3,4.5)--(5,4.5)--(5,1.5)node[below]{$Q_2$};
  \node at (0.5,3.0){$\epsilon_1$};
  \node at (0.5,2.5){$c_1$};
  \node at (5.5,3.0){$\epsilon_2$};
  \node at (5.5,2.5){$c_2$};
  \end{tikzpicture}
    \caption{A two pathway model of an espresso coffee bed. The coffee bed is divided laterally into two parts of equal volume and cross-sectional area. The total flow, $Q$, is imposed and flows through the individual beds, $Q_1$ and $Q_2$, are determined by their permeabilities. }
    \label{fig:two_pathway_model}
\end{figure}

We develop a mathematical model to investigate the feasibility of this mechanism. The simplest possible model that could show this behaviour would require two possible pathways along which flow and extraction could be different as shown in Figure~\ref{fig:two_pathway_model}. For simplicity we neglect the vertical stratification of the coffee bed used in previous models. Instead we assume that porosity and concentration of dissolved coffee are constants within each coffee bed.  

To calculate flow through the system we use the Kozeny–Carman equation for the permeability of a porous medium of spherical particles~\cite{holdich2020fundamentals,sabet2019extended} to relate porosity to permeability. We assume a constant net flow of fluid through the system, with the flow resistance of each pathway determining the relative amounts of fluid following each path. Thus flow through each pathway is given by 
\begin{align}
    Q_{1,2}&=\dfrac{Q\kappa_{1,2}}{\kappa_1+\kappa_2}\\
    \kappa_{1,2}&=\dfrac{\epsilon_{1,2}^3}{1-\epsilon_{1,2}^2}
\end{align}
where $\kappa_i$ is a dimensionless permeability of bed $i$ and $\epsilon_i$ is the porosity of bed $i$ ($i=1,2$). The volumetric flow rate, $Q$, assumed constant, is estimated as $M_\text{shot}/\rho_\text{w} t_\text{shot}$ where $M_\text{shot}$ is the mass of the shot, $\rho_\text{w}$ is the density of water, and $t_\text{shot}$ is the time taken to pour the shot. (We follow Ref.~\onlinecite{Moroney2015} in assuming that the fluid volume is unaffected by dissolved coffee content.)

Extraction is modelled following Moroney~\cite{Moroney2015}. The transfer term is given by $DS(c_\text{sat}-c)/\lambda$ where $S$ is the surface area of grains, $D$ is the diffusivity, $c_\text{sat}$ is the concentration at the surface of the grains, assumed to be saturated and $c$ is the mass concentration of dissolved coffee solids and $\lambda$ is a diffusion length. Using the Kozeny–Carman equation for permeability and using a single term to describe transfer between the solid and the liquid implicitly assumes that the distribution of ground coffee is unimodal. In fact coffee grain size distributions are better modelled as bimodal, but we follow Ref.~\onlinecite{smith2021brewing} in only considering a single transfer term for simplicity. To relate extraction to changes in porosity we assume that the soluble and insoluble components of coffee have the same density. This also allows us to relate extraction yield to porosity 
\begin{equation}
EY(t)=\dfrac{\epsilon(t)-\epsilon(0)}{1-\epsilon(0)},    
\end{equation}
where $EY(t)$ is extraction yield at time $t$ and $\epsilon(t)$ is the porosity. Since coffee grains are only partially soluble, there is an upper limit, $EY_\text{max}$, to the extraction yield.

From the considerations above the differential equations describing conservation of coffee in the solid grains and in the liquid are
\begin{align}
    \dfrac{\text{d}}{\text{d}t} \rnd{\dfrac{(1-\epsilon_i)AL\rhoc}{2}}
    &= -\dfrac{D S(\csat-c_i)}{2\lambda}\, \theta\!\left(EY_\text{max}-EY_i\right)\\
    \dfrac{\text{d}}{\text{d}t} \rnd{\dfrac{\epsilon_iALc_i}{2}}
    &=\dfrac{D S(\csat-c_i)}{2\lambda}\,\theta\!\left(EY_\text{max}-EY_i\right)-Q_ic_i
\end{align}
Where $A$ is the total area and $L$ is the thickness  of the coffee beds, $\theta(.)$ is the Heaviside function. Initial conditions are $c_{1,2}=0$ and $\epsilon_{1,2}=\epsilon_0\pm\delta$. $\rhoc$ is the mass of coffee per unit volume of the grains. Note that because the grains themselves are porous\cite{Moroney2015} this is not the same as the density of solid coffee but includes a correction of the internal porosity of the grains.

The equations can be rearranged and nondimensionalised to the form 
\begin{align}
      \dfrac{\text{d} \epsilon_i}{\text{d} \tau} &=
         \left(1-C_i\right)\,\theta\!\left(EY_\text{max}-EY_i\right) \\
      \label{nondim_c_eq}
      \alpha \epsilon_i \dfrac{\text{d}C_i}{\text{d} \tau}&=
      (1-C_i)(1-\alpha C_i)\,\theta\!\left(EY_\text{max}-EY_i\right)-\dfrac{2\beta\kappa_i}{\kappa_1+\kappa_2}C_i
\end{align}
where $c_i=\csat C_i$ where $C_i$ is a dimensionless concentration, $t=\tfrac{A L \rhoc \lambda}{D S \csat}\tau$ where $\tau$ is a dimensionless time and  the dimensionless parameters $\alpha$ and $\beta$ are given by
\begin{align}
    \alpha&=\dfrac{\csat}{\rhoc}\\
    \beta &=\dfrac{Q \lambda}{D S}.
\end{align}

\deleted{Flow instabilities} \added{Similar phenomena} in reacting porous media have been studied using continuum~\cite{jones2018reaction} and microcontinuum methods~\cite{soulaine2017mineral,soulaine2018pore}. In both approaches \deleted{instability} \added{the model developed} is investigated by direct numerical simulation\added{, as we do here}. We thus include a small difference $\delta$ in the porosities of the coffee beds in pathways 1 and 2, taking $\epsilon_{1,2}=\epsilon_0\pm\delta$ at $t=0$. \deleted{To see if there is an instability we} \added{We} simulate the evolution of the system and observe whether the difference in porosity grows or shrinks over time. To solve the differential equations we use a 4th order Runge-Kutta method with adaptive time stepping~\cite{shampine1997matlab}. To determine the parameters $D/\lambda$, $\delta$ and $EY_\text{max}$, taken to be constants independent of the grind size setting, we use least squares fitting to the data. To do this we form a sum of squares
\begin{equation}
\chi^2{(D/\lambda,\delta,EY_\text{max})}=\sum_i (EY_{\text{model},i}-EY_{\text{data},i})^2,
\end{equation}
where the summation is over different grind settings, $i$, and  $EY_{\text{data},i}$ is extraction yield data taken from Ref.~\onlinecite{cameron2020}.
Constrained BFGS~\cite{press2007numerical} minimisation is used to find optimal values of the parameters.

\section{Results}




\begin{figure}
    \centering
\begin{tikzpicture}[domain=0:3]
\begin{axis}[xlabel={$g$}, ylabel={$t_{\text{shot}}$ (s)}]
\addplot[only marks, color=black] coordinates {
 (1.100000 , 37.600000)
(1.300000 , 35.700000)
(1.500000 , 33.900000)
(1.700000 , 30.500000)
(1.900000 , 28.500000)
(2.100000 , 26.600000)
(2.300000 , 24.000000)
};
\addplot[color=black] coordinates {
 (1.000000 , 39.021429)
(1.155556 , 37.232540)
(1.311111 , 35.443651)
(1.466667 , 33.654762)
(1.622222 , 31.865873)
(1.777778 , 30.076984)
(1.933333 , 28.288095)
(2.088889 , 26.499206)
(2.244444 , 24.710317)
(2.400000 , 22.921429)
};
\node[left] at (axis cs: 2.4,38) {$t_\text{shot}=a_0+a_1g$};
\node[left] at (axis cs: 2.4,36) {$a_0=50.5\;\text{s}$};
\node[left] at (axis cs: 2.4,34) {$a_1=-11.5\;\text{s}$};
\end{axis}
\end{tikzpicture}
    \caption{Linear fit to $t_\text{shot}$ the time taken to pour a shot vs the grind size setting $g$.}
    \label{fig:t_shot_fit}
\end{figure}
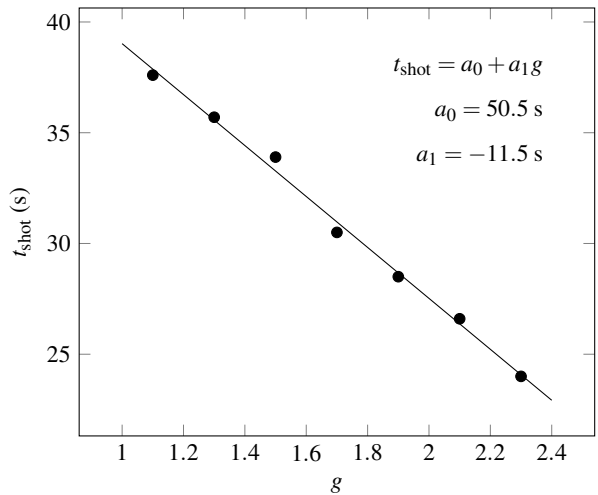

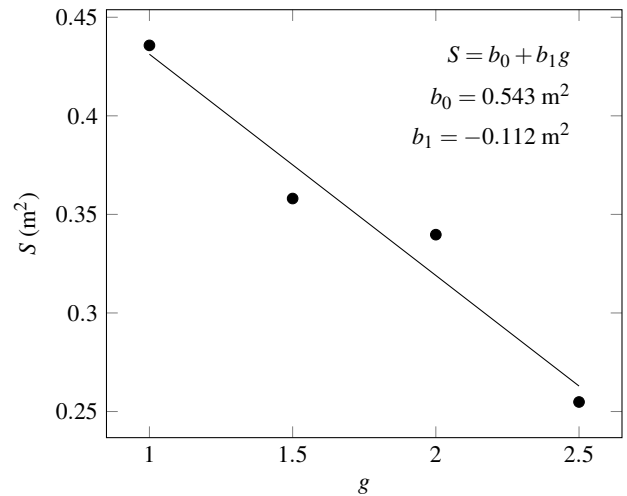
\begin{figure}
    \centering
    \begin{tikzpicture}[domain=0:3]
\begin{axis}[xlabel={$g$}, ylabel={$S$ (m$^2$)}]
\addplot[only marks, color=black] coordinates {
 (1.000000 , 0.435738)
(1.500000 , 0.358047)
(2.000000 , 0.339714)
(2.500000 , 0.254836)
};
\addplot[color=black] coordinates {
 (1.000000 , 0.431240)
(1.166667 , 0.412538)
(1.333333 , 0.393837)
(1.500000 , 0.375136)
(1.666667 , 0.356434)
(1.833333 , 0.337733)
(2.000000 , 0.319032)
(2.166667 , 0.300331)
(2.333333 , 0.281629)
(2.500000 , 0.262928)
};
\node[left] at (axis cs: 2.5,0.43) {$S=b_0+b_1g$};
\node[left] at (axis cs: 2.5,0.41) {$b_0=0.543\;\text{m}^2$};
\node[left] at (axis cs: 2.5,0.39) {$b_1=-0.112\;\text{m}^2$};
\end{axis}
\end{tikzpicture}
    \caption{Linear fit to $S$ the surface area of grains in the coffee bed vs the grind size setting $g$.}
    \label{fig:S_fit}
\end{figure}

\begin{figure}
    \centering
\begin{tikzpicture}[domain=0:3]
\begin{axis}[xlabel={$g$}, xmax=2.5]
\addplot[color=gray] coordinates {
 (1.100000 , 0.315969)
(1.300000 , 0.355387)
(1.500000 , 0.402684)
(1.700000 , 0.460114)
(1.900000 , 0.530798)
(2.100000 , 0.619165)
(2.300000 , 0.731670)
};
\addplot[color=black] coordinates {
 (1.100000 , 1.080076)
(1.300000 , 0.960277)
(1.500000 , 0.847488)
(1.700000 , 0.741708)
(1.900000 , 0.642938)
(2.100000 , 0.551178)
(2.300000 , 0.466426)
};
\node[right] at (axis cs: 2.300000,0.731670) {$\beta$};
\node[right] at (axis cs: 2.300000,0.466426) {$\tau_\text{shot}$};
\end{axis}
\end{tikzpicture}    
    \caption{Dimensionless quantities $\tau_\text{shot}$ and $\beta$ as a function of grind size setting, $g$.}
    \label{fig:tau_shot_and_beta}
\end{figure}
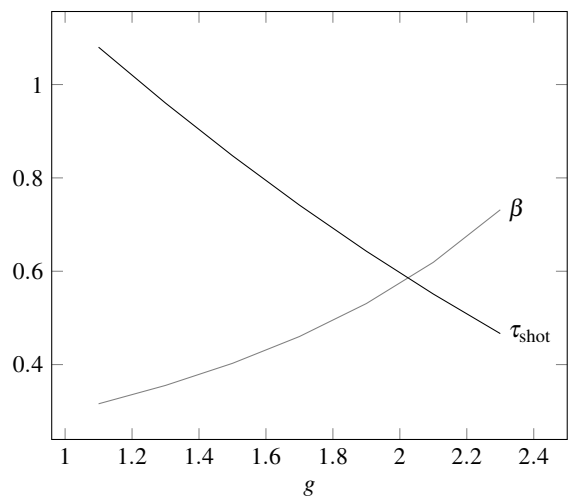

To find $Q$ we use the total volume of the shot and the time it takes to 
pour a shot, $t_\text{shot}$, assuming a uniform flow rate.
We find an expression for $t_\text{shot}$ by a linear fit to data in Ref~\onlinecite{cameron2020}. Figure~\ref{fig:t_shot_fit} shows a good fit to the data. Similarly the surface area is found by combining data on surface areas of fines and boulders in Ref~\onlinecite{cameron2020} and then performing linear regression against grind size as shown in Figure~\ref{fig:S_fit}. 
The parameters used in this model including those determined by fitting are shown in Table~\ref{tab:parameters}. The dimensionless quantities $\tau_\text{shot}$ and $\beta$ are functions of grind size setting as shown in Figure~\ref{fig:tau_shot_and_beta}. 
\begin{table}
    \centering
    \caption{Parameter values used in simulations and their sources. In the case of $\rhoc$ both the value from the literature and the unphysical value used here are reported.}
    \label{tab:parameters}
\begin{tabular}{l r l l}
\hline
Parameter & Value & & Source\\ 
\hline\hline
$M_\text{shot}$ &    0.04 & $\text{kg}$& Ref.~\onlinecite{cameron2020} \\ 
$\rho_\text{w}$ &  997 & $\text{kg}\,\text{m}^{-3}$ &Ref.~\onlinecite{cameron2020} \\ 
$\epsilon_0$ &       0.173 && Ref.~\onlinecite{cameron2020} \\ 
$c_\text{sat}$ &   212.4 & $\text{kg}\,\text{m}^{-3}$ & Ref.~\onlinecite{cameron2020}\\ 
$\rho_\text{c}$ &  399 & $\text{kg}\,\text{m}^{-3}$ & Ref.~\onlinecite{cameron2020} \\ 
$\rho_\text{c}$ &  798 & $\text{kg}\,\text{m}^{-3}$ & Imposed \\ 
$\alpha$ &           3.76 && Calculated \\ 
$\lambda/D$ &        0.125$\times 10^6$ & $\text{s}\,\text{m}^{-1}$ & Best fit\\ 
$\delta$ &           0.035    && Best fit \\ 
$EY_\text{max}$ &   33.8   & $\%$ & Best fit\\ 
\hline
\end{tabular}
\end{table}


\begin{figure}
    \centering
\begin{tikzpicture}[domain=0:3]
\begin{axis}[xlabel={$g$}, ylabel={Extraction Yield}]
\addplot[only marks, color=black] coordinates {
 (1.100000 , 21.300000)
(1.300000 , 22.400000)
(1.500000 , 22.500000)
(1.700000 , 23.000000)
(1.900000 , 22.500000)
(2.100000 , 21.500000)
(2.300000 , 20.600000)
};
\addplot[color=black] coordinates {
 (1.100000 , 22.099607)
(1.300000 , 22.178030)
(1.500000 , 22.262789)
(1.700000 , 22.346602)
(1.900000 , 22.421323)
(2.100000 , 21.658611)
(2.300000 , 20.385476)
};
\end{axis}
\end{tikzpicture}
    \caption{Points show experimental data for Extraction yield, $EY$ as a function of grind size setting, $g$. The line shows the overall extraction yield from the best fit two pathway model. The model reproduces the qualitative trend. }
    \label{fig:EY_fit}
\end{figure}
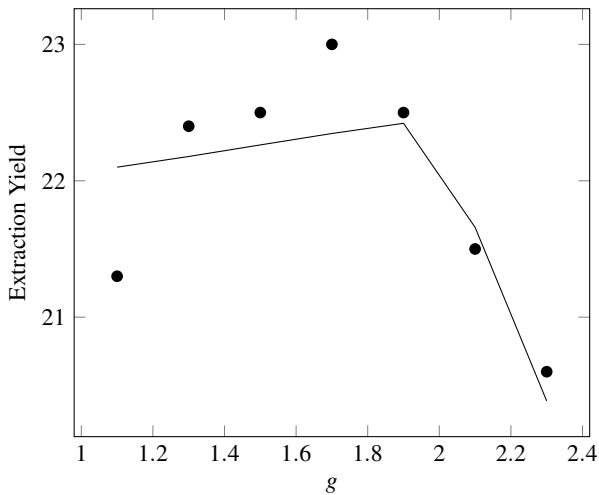

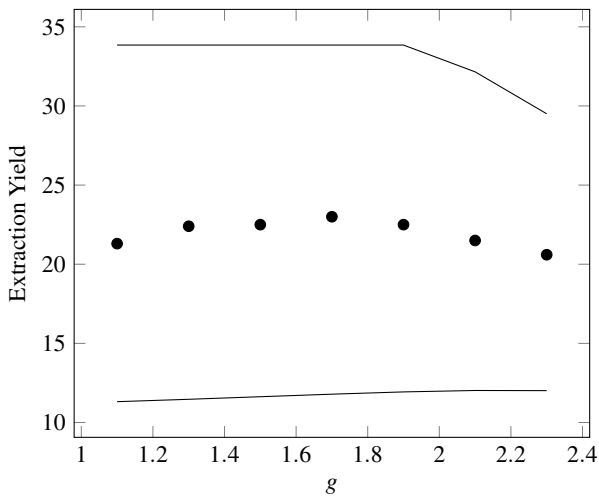
\begin{figure}
    \centering
\begin{tikzpicture}[domain=0:3]
\begin{axis}[xlabel={$g$}, ylabel={Extraction Yield}]
\addplot[only marks, color=black] coordinates {
 (1.100000 , 21.300000)
(1.300000 , 22.400000)
(1.500000 , 22.500000)
(1.700000 , 23.000000)
(1.900000 , 22.500000)
(2.100000 , 21.500000)
(2.300000 , 20.600000)
};
\addplot[color=black] coordinates {
 (1.100000 , 33.846678)
(1.300000 , 33.846678)
(1.500000 , 33.846678)
(1.700000 , 33.846678)
(1.900000 , 33.846678)
(2.100000 , 32.155432)
(2.300000 , 29.505408)
};
\addplot[color=black] coordinates {
 (1.100000 , 11.316527)
(1.300000 , 11.466936)
(1.500000 , 11.629499)
(1.700000 , 11.790248)
(1.900000 , 11.933557)
(2.100000 , 12.023182)
(2.300000 , 12.013946)
};
\end{axis}
\end{tikzpicture}
    \caption{Points show experimental data for Extraction yield, $EY$ as a function of grind size setting, $g$. The two lines show the extraction yield of the individual pathways from the best fit two pathway model.}
    \label{fig:EYs_fit}
\end{figure}

Plotting the best fit model against the data, as seen in Figure~\ref{fig:EY_fit}, shows that although the two pathway model is very simple it can reproduce the qualitative features of the data, including decreasing extraction yield with decreasing grind size. Figure~\ref{fig:EYs_fit} shows the extraction yield from the individual pathways. The plot suggests that the transition from the expected behaviour of extraction increasing as grind size decreased to the surprising behaviour of decreasing extraction with decreasing grind size corresponds to the onset of saturation in one of the pathways.  

\begin{figure}
    \centering
    \begin{tikzpicture}[domain=0:3]
\begin{axis}[xlabel={$\tau$}, xmax=1.25]
\addplot[color=black] coordinates {
 (0.000000 , 0.208193)
(0.021602 , 0.226397)
(0.042790 , 0.239775)
(0.064391 , 0.250774)
(0.085993 , 0.260129)
(0.107594 , 0.268432)
(0.129196 , 0.276043)
(0.150797 , 0.283189)
(0.172399 , 0.290018)
(0.194000 , 0.296628)
(0.215602 , 0.303084)
(0.237203 , 0.309432)
(0.258805 , 0.315704)
(0.280406 , 0.321920)
(0.302008 , 0.328096)
(0.323610 , 0.334243)
(0.345211 , 0.340369)
(0.366813 , 0.346480)
(0.388414 , 0.352578)
(0.410016 , 0.358668)
(0.431617 , 0.364752)
(0.453219 , 0.370831)
(0.474820 , 0.376905)
(0.496422 , 0.382978)
(0.518023 , 0.389047)
(0.539625 , 0.395116)
(0.561226 , 0.401182)
(0.582828 , 0.407248)
(0.604429 , 0.413313)
(0.626031 , 0.419378)
(0.647632 , 0.425442)
(0.669234 , 0.431505)
(0.690835 , 0.437568)
(0.712437 , 0.443631)
(0.734038 , 0.449694)
(0.755640 , 0.455756)
(0.777241 , 0.461819)
(0.798843 , 0.467881)
(0.820444 , 0.473943)
(0.842046 , 0.477219)
(0.863647 , 0.477219)
(0.885249 , 0.477219)
(0.906850 , 0.477219)
(0.928452 , 0.477219)
(0.950053 , 0.477219)
(0.971655 , 0.477219)
(0.993257 , 0.477219)
(1.014858 , 0.477219)
(1.036460 , 0.477219)
(1.058061 , 0.477219)
(1.079663 , 0.477219)
(1.080076 , 0.477219)
};
\addplot[color=black] coordinates {
 (0.000000 , 0.137407)
(0.021602 , 0.154140)
(0.042790 , 0.164644)
(0.064391 , 0.172154)
(0.085993 , 0.177762)
(0.107594 , 0.182180)
(0.129196 , 0.185818)
(0.150797 , 0.188925)
(0.172399 , 0.191656)
(0.194000 , 0.194112)
(0.215602 , 0.196361)
(0.237203 , 0.198447)
(0.258805 , 0.200401)
(0.280406 , 0.202244)
(0.302008 , 0.203991)
(0.323610 , 0.205652)
(0.345211 , 0.207237)
(0.366813 , 0.208752)
(0.388414 , 0.210200)
(0.410016 , 0.211586)
(0.431617 , 0.212913)
(0.453219 , 0.214184)
(0.474820 , 0.215401)
(0.496422 , 0.216567)
(0.518023 , 0.217682)
(0.539625 , 0.218751)
(0.561226 , 0.219773)
(0.582828 , 0.220750)
(0.604429 , 0.221685)
(0.626031 , 0.222579)
(0.647632 , 0.223432)
(0.669234 , 0.224248)
(0.690835 , 0.225027)
(0.712437 , 0.225770)
(0.734038 , 0.226479)
(0.755640 , 0.227156)
(0.777241 , 0.227800)
(0.798843 , 0.228415)
(0.820444 , 0.229001)
(0.842046 , 0.229558)
(0.863647 , 0.230092)
(0.885249 , 0.230609)
(0.906850 , 0.231114)
(0.928452 , 0.231610)
(0.950053 , 0.232100)
(0.971655 , 0.232587)
(0.993257 , 0.233071)
(1.014858 , 0.233556)
(1.036460 , 0.234040)
(1.058061 , 0.234526)
(1.079663 , 0.235013)
(1.080076 , 0.235022)
};
\addplot[color=gray] coordinates {
 (0.000000 , 0.000000)
(0.021602 , 0.284209)
(0.042790 , 0.439324)
(0.064391 , 0.534688)
(0.085993 , 0.594738)
(0.107594 , 0.633802)
(0.129196 , 0.659813)
(0.150797 , 0.677441)
(0.172399 , 0.689557)
(0.194000 , 0.697984)
(0.215602 , 0.703903)
(0.237203 , 0.708100)
(0.258805 , 0.711098)
(0.280406 , 0.713258)
(0.302008 , 0.714824)
(0.323610 , 0.715967)
(0.345211 , 0.716807)
(0.366813 , 0.717429)
(0.388414 , 0.717891)
(0.410016 , 0.718237)
(0.431617 , 0.718498)
(0.453219 , 0.718695)
(0.474820 , 0.718845)
(0.496422 , 0.718961)
(0.518023 , 0.719049)
(0.539625 , 0.719118)
(0.561226 , 0.719172)
(0.582828 , 0.719214)
(0.604429 , 0.719247)
(0.626031 , 0.719274)
(0.647632 , 0.719295)
(0.669234 , 0.719311)
(0.690835 , 0.719325)
(0.712437 , 0.719335)
(0.734038 , 0.719344)
(0.755640 , 0.719351)
(0.777241 , 0.719356)
(0.798843 , 0.719361)
(0.820444 , 0.719364)
(0.842046 , 0.701838)
(0.863647 , 0.665196)
(0.885249 , 0.630471)
(0.906850 , 0.597562)
(0.928452 , 0.566376)
(0.950053 , 0.536821)
(0.971655 , 0.508812)
(0.993257 , 0.482270)
(1.014858 , 0.457116)
(1.036460 , 0.433278)
(1.058061 , 0.410688)
(1.079663 , 0.389279)
(1.080076 , 0.388881)
};
\addplot[color=gray] coordinates {
 (0.000000 , 0.000000)
(0.021602 , 0.397606)
(0.042790 , 0.591345)
(0.064391 , 0.703540)
(0.085993 , 0.771955)
(0.107594 , 0.815928)
(0.129196 , 0.845380)
(0.150797 , 0.865834)
(0.172399 , 0.880542)
(0.194000 , 0.891494)
(0.215602 , 0.899941)
(0.237203 , 0.906689)
(0.258805 , 0.912263)
(0.280406 , 0.917014)
(0.302008 , 0.921174)
(0.323610 , 0.924903)
(0.345211 , 0.928308)
(0.366813 , 0.931465)
(0.388414 , 0.934423)
(0.410016 , 0.937221)
(0.431617 , 0.939882)
(0.453219 , 0.942426)
(0.474820 , 0.944865)
(0.496422 , 0.947208)
(0.518023 , 0.949463)
(0.539625 , 0.951635)
(0.561226 , 0.953728)
(0.582828 , 0.955746)
(0.604429 , 0.957690)
(0.626031 , 0.959564)
(0.647632 , 0.961369)
(0.669234 , 0.963107)
(0.690835 , 0.964781)
(0.712437 , 0.966391)
(0.734038 , 0.967939)
(0.755640 , 0.969428)
(0.777241 , 0.970858)
(0.798843 , 0.972232)
(0.820444 , 0.973550)
(0.842046 , 0.974784)
(0.863647 , 0.975719)
(0.885249 , 0.976392)
(0.906850 , 0.976867)
(0.928452 , 0.977191)
(0.950053 , 0.977400)
(0.971655 , 0.977523)
(0.993257 , 0.977579)
(1.014858 , 0.977584)
(1.036460 , 0.977550)
(1.058061 , 0.977485)
(1.079663 , 0.977398)
(1.080076 , 0.977396)
};
\node[right] at (axis cs: 1.080076,0.477219) {$\epsilon_1$};
\node[right] at (axis cs: 1.080076,0.235022) {$\epsilon_2$};
\node[right] at (axis cs: 1.080076,0.388881) {$C_1$};
\node[right] at (axis cs: 1.080076,0.977396) {$C_2$};
\node[right] at (axis cs: 0,0.977396) {$g=1.1$};
\end{axis}
\end{tikzpicture}
    \caption{Porosity and dimensionless concentration of dissolved coffee solids as a function of dimensionless time for finely ground coffee. All the soluble coffee is extracted from bed~1 at $\tau\approx0.8$. }
    \label{fig:fine_grains}
\end{figure}
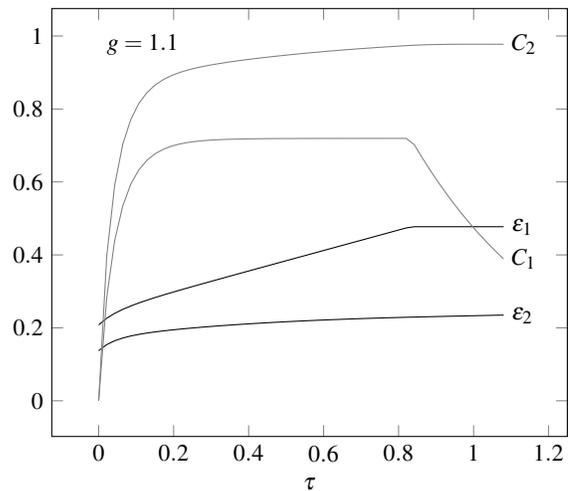

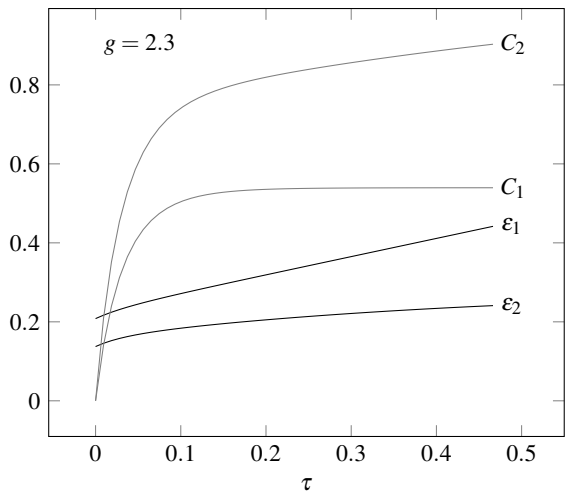
\begin{figure}
    \centering
\begin{tikzpicture}[domain=0:3]
\begin{axis}[xlabel={$\tau$}, xmax=0.55]
\addplot[color=black] coordinates {
 (0.000000 , 0.208193)
(0.009329 , 0.216826)
(0.018657 , 0.224352)
(0.027986 , 0.231090)
(0.037314 , 0.237252)
(0.046643 , 0.242987)
(0.055971 , 0.248402)
(0.065300 , 0.253573)
(0.074628 , 0.258558)
(0.083957 , 0.263400)
(0.093285 , 0.268129)
(0.102614 , 0.272771)
(0.111942 , 0.277345)
(0.121271 , 0.281865)
(0.130599 , 0.286341)
(0.139928 , 0.290784)
(0.149256 , 0.295199)
(0.158585 , 0.299592)
(0.167913 , 0.303968)
(0.177242 , 0.308329)
(0.186571 , 0.312679)
(0.195899 , 0.317020)
(0.205228 , 0.321352)
(0.214556 , 0.325679)
(0.223885 , 0.330000)
(0.233213 , 0.334317)
(0.242542 , 0.338631)
(0.251870 , 0.342942)
(0.261199 , 0.347250)
(0.270527 , 0.351556)
(0.279856 , 0.355861)
(0.289184 , 0.360164)
(0.298513 , 0.364466)
(0.307841 , 0.368767)
(0.317170 , 0.373068)
(0.326498 , 0.377367)
(0.335827 , 0.381666)
(0.345155 , 0.385964)
(0.354484 , 0.390262)
(0.363813 , 0.394560)
(0.373141 , 0.398857)
(0.382470 , 0.403154)
(0.391798 , 0.407451)
(0.401127 , 0.411747)
(0.410455 , 0.416044)
(0.419784 , 0.420340)
(0.429112 , 0.424636)
(0.438441 , 0.428932)
(0.447769 , 0.433228)
(0.457098 , 0.437523)
(0.466426 , 0.441819)
(0.466426 , 0.441819)
};
\addplot[color=black] coordinates {
 (0.000000 , 0.137407)
(0.009329 , 0.145696)
(0.018657 , 0.152370)
(0.027986 , 0.157915)
(0.037314 , 0.162641)
(0.046643 , 0.166757)
(0.055971 , 0.170408)
(0.065300 , 0.173698)
(0.074628 , 0.176706)
(0.083957 , 0.179487)
(0.093285 , 0.182084)
(0.102614 , 0.184531)
(0.111942 , 0.186852)
(0.121271 , 0.189068)
(0.130599 , 0.191193)
(0.139928 , 0.193240)
(0.149256 , 0.195218)
(0.158585 , 0.197134)
(0.167913 , 0.198996)
(0.177242 , 0.200808)
(0.186571 , 0.202573)
(0.195899 , 0.204296)
(0.205228 , 0.205979)
(0.214556 , 0.207624)
(0.223885 , 0.209232)
(0.233213 , 0.210806)
(0.242542 , 0.212347)
(0.251870 , 0.213856)
(0.261199 , 0.215334)
(0.270527 , 0.216781)
(0.279856 , 0.218199)
(0.289184 , 0.219587)
(0.298513 , 0.220947)
(0.307841 , 0.222280)
(0.317170 , 0.223584)
(0.326498 , 0.224862)
(0.335827 , 0.226113)
(0.345155 , 0.227338)
(0.354484 , 0.228538)
(0.363813 , 0.229712)
(0.373141 , 0.230860)
(0.382470 , 0.231984)
(0.391798 , 0.233084)
(0.401127 , 0.234159)
(0.410455 , 0.235211)
(0.419784 , 0.236239)
(0.429112 , 0.237244)
(0.438441 , 0.238226)
(0.447769 , 0.239185)
(0.457098 , 0.240123)
(0.466426 , 0.241038)
(0.466426 , 0.241038)
};
\addplot[color=gray] coordinates {
 (0.000000 , 0.000000)
(0.009329 , 0.140745)
(0.018657 , 0.239995)
(0.027986 , 0.311727)
(0.037314 , 0.364535)
(0.046643 , 0.403974)
(0.055971 , 0.433772)
(0.065300 , 0.456501)
(0.074628 , 0.473980)
(0.083957 , 0.487516)
(0.093285 , 0.498063)
(0.102614 , 0.506328)
(0.111942 , 0.512836)
(0.121271 , 0.517986)
(0.130599 , 0.522077)
(0.139928 , 0.525342)
(0.149256 , 0.527957)
(0.158585 , 0.530059)
(0.167913 , 0.531755)
(0.177242 , 0.533127)
(0.186571 , 0.534242)
(0.195899 , 0.535150)
(0.205228 , 0.535892)
(0.214556 , 0.536500)
(0.223885 , 0.537000)
(0.233213 , 0.537412)
(0.242542 , 0.537753)
(0.251870 , 0.538036)
(0.261199 , 0.538271)
(0.270527 , 0.538467)
(0.279856 , 0.538631)
(0.289184 , 0.538769)
(0.298513 , 0.538884)
(0.307841 , 0.538982)
(0.317170 , 0.539065)
(0.326498 , 0.539135)
(0.335827 , 0.539194)
(0.345155 , 0.539245)
(0.354484 , 0.539289)
(0.363813 , 0.539326)
(0.373141 , 0.539359)
(0.382470 , 0.539387)
(0.391798 , 0.539411)
(0.401127 , 0.539432)
(0.410455 , 0.539450)
(0.419784 , 0.539466)
(0.429112 , 0.539480)
(0.438441 , 0.539493)
(0.447769 , 0.539504)
(0.457098 , 0.539514)
(0.466426 , 0.539522)
(0.466426 , 0.539522)
};
\addplot[color=gray] coordinates {
 (0.000000 , 0.000000)
(0.009329 , 0.208809)
(0.018657 , 0.351811)
(0.027986 , 0.453961)
(0.037314 , 0.529184)
(0.046643 , 0.585901)
(0.055971 , 0.629500)
(0.065300 , 0.663579)
(0.074628 , 0.690622)
(0.083957 , 0.712388)
(0.093285 , 0.730147)
(0.102614 , 0.744832)
(0.111942 , 0.757140)
(0.121271 , 0.767594)
(0.130599 , 0.776591)
(0.139928 , 0.784438)
(0.149256 , 0.791371)
(0.158585 , 0.797572)
(0.167913 , 0.803186)
(0.177242 , 0.808324)
(0.186571 , 0.813076)
(0.195899 , 0.817512)
(0.205228 , 0.821688)
(0.214556 , 0.825648)
(0.223885 , 0.829428)
(0.233213 , 0.833058)
(0.242542 , 0.836559)
(0.251870 , 0.839950)
(0.261199 , 0.843248)
(0.270527 , 0.846462)
(0.279856 , 0.849604)
(0.289184 , 0.852681)
(0.298513 , 0.855699)
(0.307841 , 0.858665)
(0.317170 , 0.861581)
(0.326498 , 0.864451)
(0.335827 , 0.867278)
(0.345155 , 0.870065)
(0.354484 , 0.872812)
(0.363813 , 0.875522)
(0.373141 , 0.878195)
(0.382470 , 0.880832)
(0.391798 , 0.883435)
(0.401127 , 0.886002)
(0.410455 , 0.888535)
(0.419784 , 0.891034)
(0.429112 , 0.893499)
(0.438441 , 0.895931)
(0.447769 , 0.898329)
(0.457098 , 0.900693)
(0.466426 , 0.903023)
(0.466426 , 0.903023)
};
\node[right] at (axis cs: 0.466426,0.441819) {$\epsilon_1$};
\node[right] at (axis cs: 0.466426,0.241038) {$\epsilon_2$};
\node[right] at (axis cs: 0.466426,0.539522) {$C_1$};
\node[right] at (axis cs: 0.466426,0.903023) {$C_2$};
\node[right] at (axis cs: 0,0.903023) {$g=2.3$};
\end{axis}
\end{tikzpicture}
    \caption{Porosity and dimensionless concentration of dissolved coffee solids as a function of dimensionless time for coarsely  ground coffee.}
    \label{fig:coarse_grains}
\end{figure}

\begin{figure}
\begin{tikzpicture}[domain=0:3]
\begin{axis}[xlabel={$\tau$}, xmax=0.536390]
\addplot[color=black] coordinates {
 (0.000000 , 0.781053)
(0.009329 , 0.771960)
(0.018657 , 0.766538)
(0.027986 , 0.763483)
(0.037314 , 0.762033)
(0.046643 , 0.761710)
(0.055971 , 0.762201)
(0.065300 , 0.763293)
(0.074628 , 0.764835)
(0.083957 , 0.766717)
(0.093285 , 0.768860)
(0.102614 , 0.771203)
(0.111942 , 0.773701)
(0.121271 , 0.776319)
(0.130599 , 0.779028)
(0.139928 , 0.781808)
(0.149256 , 0.784642)
(0.158585 , 0.787516)
(0.167913 , 0.790419)
(0.177242 , 0.793343)
(0.186571 , 0.796281)
(0.195899 , 0.799226)
(0.205228 , 0.802174)
(0.214556 , 0.805121)
(0.223885 , 0.808064)
(0.233213 , 0.810999)
(0.242542 , 0.813925)
(0.251870 , 0.816839)
(0.261199 , 0.819740)
(0.270527 , 0.822625)
(0.279856 , 0.825495)
(0.289184 , 0.828346)
(0.298513 , 0.831179)
(0.307841 , 0.833992)
(0.317170 , 0.836785)
(0.326498 , 0.839556)
(0.335827 , 0.842306)
(0.345155 , 0.845032)
(0.354484 , 0.847735)
(0.363813 , 0.850414)
(0.373141 , 0.853068)
(0.382470 , 0.855697)
(0.391798 , 0.858301)
(0.401127 , 0.860879)
(0.410455 , 0.863430)
(0.419784 , 0.865955)
(0.429112 , 0.868452)
(0.438441 , 0.870921)
(0.447769 , 0.873363)
(0.457098 , 0.875777)
(0.466426 , 0.878163)
(0.466426 , 0.878163)
};
\addplot[color=black] coordinates {
 (0.000000 , 0.218947)
(0.009329 , 0.228040)
(0.018657 , 0.233462)
(0.027986 , 0.236517)
(0.037314 , 0.237967)
(0.046643 , 0.238290)
(0.055971 , 0.237799)
(0.065300 , 0.236707)
(0.074628 , 0.235165)
(0.083957 , 0.233283)
(0.093285 , 0.231140)
(0.102614 , 0.228797)
(0.111942 , 0.226299)
(0.121271 , 0.223681)
(0.130599 , 0.220972)
(0.139928 , 0.218192)
(0.149256 , 0.215358)
(0.158585 , 0.212484)
(0.167913 , 0.209581)
(0.177242 , 0.206657)
(0.186571 , 0.203719)
(0.195899 , 0.200774)
(0.205228 , 0.197826)
(0.214556 , 0.194879)
(0.223885 , 0.191936)
(0.233213 , 0.189001)
(0.242542 , 0.186075)
(0.251870 , 0.183161)
(0.261199 , 0.180260)
(0.270527 , 0.177375)
(0.279856 , 0.174505)
(0.289184 , 0.171654)
(0.298513 , 0.168821)
(0.307841 , 0.166008)
(0.317170 , 0.163215)
(0.326498 , 0.160444)
(0.335827 , 0.157694)
(0.345155 , 0.154968)
(0.354484 , 0.152265)
(0.363813 , 0.149586)
(0.373141 , 0.146932)
(0.382470 , 0.144303)
(0.391798 , 0.141699)
(0.401127 , 0.139121)
(0.410455 , 0.136570)
(0.419784 , 0.134045)
(0.429112 , 0.131548)
(0.438441 , 0.129079)
(0.447769 , 0.126637)
(0.457098 , 0.124223)
(0.466426 , 0.121837)
(0.466426 , 0.121837)
};
\node[below] at (axis cs: 0.466426,0.878163) {$Q_1/Q$};
\node[above] at (axis cs: 0.466426,0.121837) {$Q_2/Q$};
\node[right] at (axis cs: 0,0.878163) {$g=2.3$};
\end{axis}
\end{tikzpicture}
    \caption{Flow rates as a function of dimensionless time for coarsely ground coffee.} 
    \label{fig:coarse_grain_flow}
\end{figure}
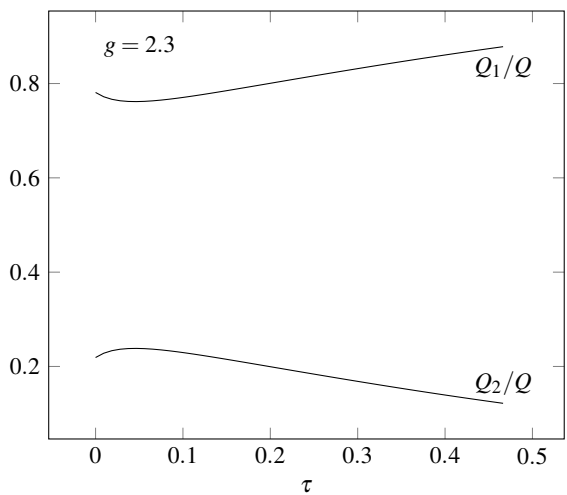

To see how extraction varies with time we show time dependent results from the two extreme grind sizes $g=1.1$ and $g=2.3$. The plot for finely ground coffee $g=1.1$, Figure~\ref{fig:fine_grains} shows clearly that one pathway hits \deleted{saturation point} \added{full extraction}. For the coarsely ground sample, $g=2.3$, Figure~\ref{fig:coarse_grains} shows \deleted{the instability} \added{an increasing difference in porosity between the two pathways}, but \added{not complete extraction of all available coffee in either pathway} \deleted{no saturation}. Figure~\ref{fig:coarse_grain_flow} shows the flow through the system. The figures suggest that a complex interplay between a number of interacting phenomena are needed to explain the observed results. The model was set up to investigate the hypothesis that a \deleted{flow-dissolution instability} \added{an interaction between flow and dissolution resulting in uneven extraction} could explain the unexpected decrease in extraction with decreasing grind size with the onset \deleted{of the instability} \added{of uneven extraction} occurring at the point where the trend changes from extraction increasing with decreasing grind size to extraction decreasing with increasing grind size. In fact the model shows that this instability is present at all grind sizes. The decreasing extraction with decreasing grind size also requires that all available coffee is extracted from one part of the coffee bed. This can be seen in Figure~\ref{fig:fine_grains}: in the more porous, faster flowing pathway, i.e. pathway 1, coffee is being extracted more quickly as can be seen by $C_1<C_2$ (lower concentration indicates faster extraction since the extraction rate is proportional to $\csat-c$). At $\tau\approx 0.8$ all the available coffee has been dissolved from the grains in pathway 1 as can be seen from $\epsilon_1$ remaining constant (the porosity no longer changes due to dissolution) and the rapid decrease in $C_1$ coffee in solution is washed out of the system by the flow but in not replaced by further dissolution.

\section{Discussion}

 The modelling predicts a maximum extraction yield of 33.8\%. This can be compared with a value calculated by Smith and Lee~\cite{smith2021brewing} from a dataset given in Ref~\onlinecite{Moroney2015} of 30.3\%. The fitted value is plausible, but perhaps a little high. In this work: $\lambda/D=0.125\times 10^6 \unit{s} \unit{m}^{-1}$. This can be compared to the values calculated from parameters given in Ref.~\onlinecite{Moroney2015}. Two different coffees, ground differently, values of   $0.128\times 10^6 \unit{s} \unit{m}^{-1}$, $0.210\times 10^6 \unit{s} \unit{m}^{-1}$ are obtained. This suggests that the fitting has produced a realistic value for this parameter. The best fit value of $\delta$ looks a little high, as it is a significant fraction of the porosity. This may be a result of the simplicity of the model.  

In Ref~\onlinecite{cameron2020} an alternative but related explanation is proposed for the observed trends, namely that at smaller grind sizes clogging appears. In other words some pathways through the coffee bed are blocked by fines. In this modelling framework this would be modelled by a large initial difference of permeability between the two pathways representing the difference between clogged and unclogged pathways. I.e.\ by a large initial value of $\delta$ since this parameter controls the initial difference in porosity and thus permeability between the two pathways of the model.  There is thus some support for this within the model. However, the same value of $\delta$ is used for all the simulations, so the model does not support the idea that the onset of clogging is responsible for the turnover in the trend. 

One unsatisfactory feature of the model is that the decreasing trend in extraction with decreasing grind size is not seen with 
the value of $\rhoc$ 
observed experimentally in Ref.~\onlinecite{cameron2020}. If this value is used then the extraction yield remains constant below the critical grind size but does not decrease. In order to see the observed behaviour of decreasing extraction with decreasing grind size an unphysical value of $\rhoc$ twice the size of the physical value is used here. (Both values for $\rhoc$ are given in Table~\ref{tab:parameters}.) Given how simple the model is, it is not surprising that some adjustments of parameters must be made to reproduce the observed behaviour. In particular  the large value of $\rhoc$ needed may indicate
that some aspect of the interplay between flow and dissolution is not being correctly captured by this simple model. Adding greater complexity, such as including a bimodal size distribution may be needed.

These results also have implications for the taste of coffee. As is shown by the coffee brewing control chart over-extraction of coffee results in an excess of bitter flavours. Thus, given that beyond the turnover point we expect at least some part of the coffee bed to be significantly over extracted, in fact for all soluble compounds to be dissolved, we would expect the coffee brewed beyond this point to have a more bitter taste than coffee brewed with the same overall extraction yield at the high grind size. 

\section{Conclusions}

The aim of this study was to investigate whether  \deleted{a flow instability} \added{an interaction between flow and dissolution leading to uneven extraction between different flow pathways} could explain the anomalous trend of extraction yield with grind size seen experimentally by Cameron et al.~\cite{cameron2020}. The proposed \deleted{instability} \added{mechanism} was a positive feedback loop between flow and extraction. We investigated this in the simplest possible model capable of showing such \deleted{an instability} \added{phenomena}, one in which there were two potential pathways for flow. 
The \deleted{instability} \added{mechanism} was investigated by simulating the evolution of the system with an initial difference in porosity between the two pathways built in. Model parameters were either taken from the original paper, determined by curve fitting or in one case chosen in order to see the desired behaviour.
Surprisingly for such a simple model the observed trend of a peak in extraction yield with grind size was reproduced. 

In building this model the initial assumption was that the peak in extraction yield would indicate the onset of \deleted{instability} \added{uneven flow between the different pathways}. In fact the model suggests that \deleted{instability} \added{uneven flow between pathways} is always present and that the peak in extraction yield is due to dissolution of all soluble coffee from one part of the coffee bed. In other words \deleted{instability is always present} \added{differences in porosities between pathways will always be amplified} but as the amount of coffee extracted from one region decreases, an overcompensating increase in the extraction of coffee from other regions masks this trend until all the soluble coffee is dissolved. This has important implications for the taste of the coffee, the non-uniform extraction suggests that the average extraction yield may not be a good guide to taste particularly at fine grind sizes. This result also has implications for simulations of coffee brewing. These have typically focussed on variations with depth but have not considered lateral variations in extraction. 

A number of extensions to the model which may result in a better fit to the observed data and may remove the need to incorporate an unphysical parameter value such as taking $\rhoc$ to be twice the size of the measured value. One example would be to include a bimodal description of the coffee grains. It is possible that the large value of $\rhoc$ is needed to compensate for the model not including the densely packed fine grains. Another feature that could be included that is already part of existing models is to include vertical stratification. The model currently assumes that porosity, and dissolved coffee concentration are uniform across the depth of the coffee bed. Relaxing this assumption would make the model more realistic as simulations show there is significant variation with depth. A final improvement to the model would be to remove the assumption that there \deleted{is} \added{are} two equally weighted pathways for flow. This could be done either by including a larger number of pathways or by allowing the areas of the pathways to be different fractions of the total cross sectional area, this split could be included as one of the fitting parameters of the model.

This work suggests that if the instability in the flow could be eliminated or reduced then the taste of espresso coffee could be improved  and significant financial and environmental costs associated with wasted material could be eliminated. One obvious thing to check first would be be to make sure that flow was as even as possible. Depending on the exact nature of the instability it may be possible to prevent it by changing the aspect ratio of the coffee bed. If there is a macroscopic sized lateral lengthscale associated with the instability then the instability could be eliminated by reducing the horizontal size of the coffee bed to make it smaller than the scale of the instability.

\appendix

\section{Ancillary Files}

\verb|coffee_odes.wxmx|: wxMaxima computer algebra system file carrying out the derivation and nondimensionalisation of the equations of the two pathway model.

\verb|parameter_fitting.m|: octave (matlab clone) file to carry out least squares parameter fitting.

\verb|latex_output.m|: octave file that generates a latex file containining the graphs and table for the paper.

\bibliography{biblio} 
\bibliographystyle{plain}

\end{document}